\documentclass[12pt,preprint]{aastex}

\shorttitle{Quasar Galaxy Hosts}
\shortauthors{Young et al.}

\begin{document}

%% LaTeX will automatically break titles if they run longer than
%% one line. However, you may use \\ to force a line break if
%% you desire.

\title{The Contribution from Scattered Light to Quasar Galaxy Hosts}

%% Use \author, \affil, and the \and command to format
%% author and affiliation information.
%% Note that \email has replaced the old \authoremail command
%% from AASTeX v4.0. You can use \email to mark an email address
%% anywhere in the paper, not just in the front matter.
%% As in the title, use \\ to force line breaks.

\author{S. Young\altaffilmark{1}, D. J. Axon\altaffilmark{1} and A. Robinson\altaffilmark{1}}
\affil{Department of Physics, Rochester Institute of Technology, 54 Lomb Memorial Drive, Rochester, NY, 14623, USA}

\and

\author{A. Capetti}
\affil{Osservatorio Astronomico di Torino, Strada Osservatorio 20, I-10025 Pino Torinese, Italy}

%% Notice that each of these authors has alternate affiliations, which
%% are identified by the \altaffilmark after each name.  Specify alternate
%% affiliation information with \altaffiltext, with one command per each
%% affiliation.

\altaffiltext{1}{Centre for Astrophysics Research, Science \& Technology Research Institute, University of Hertfordshire, Hatfield, AL10 9AB, UK}

%% Mark off your abstract in the ``abstract'' environment. In the manuscript
%% style, abstract will output a Received/Accepted line after the
%% title and affiliation information. No date will appear since the author
%% does not have this information. The dates will be filled in by the
%% editorial office after submission.

\begin{abstract}

We present models representing the scattering of quasar radiation off
free electrons and dust grains in geometries that approximate the
structure of quasar host galaxies. We show that, for reasonable
assumptions, scattering alone can easily produce ratios of nuclear
(point source) to extended fluxes comparable to those determined in
studies of quasar hosts. This result suggests that scattered quasar light, as well as stellar emission from the host galaxy, contributes significantly to the detected extended flux, leading to uncertainty in the inferred properties of quasar host. A significant contribution from scattered quasar light will
lead to overestimates of the luminosity and hence mass of the host
galaxy, and may also distort its morphology. Scattering
of quasar light within the host galaxy may provide alternative
explanations for the apparent peak in host luminosity at {\it z} = 2--3; possibly the
overall average higher luminosity of radio-loud host galaxies relative to those of
radio-quiet quasars (RQQs), and the apparent preference of high-luminosity 
RQQs for spheroidal rather than disk galaxies.

\end{abstract}

%% Keywords should appear after the \end{abstract} command. The uncommented
%% example has been keyed in ApJ style. See the instructions to authors
%% for the journal to which you are submitting your paper to determine
%% what keyword punctuation is appropriate.


\section{Introduction}

There have been numerous ground-based \citep[e.g.][]{ATB98, 
Fal08} and space-based \citep[e.g.][]{Kuk01, McL04} studies to characterize the host galaxies of quasars at {\it z} $\ga$ 0.4.  These studies suggest that 
the host galaxies of radio-loud quasars (RLQs) at high redshift are $\sim$ 2.5--3 mag brighter than at the present epoch and have evolved in a 
manner that is consistent with a formation epoch at {\it z} $\sim$ 3 and 
subsequent simple passive stellar evolution.  A similar evolutionary trend 
has been suggested for radio-quiet quasar (RQQ) hosts with the addition of a 
small increase in active galactic nucleus (AGN) fueling efficiency with increasing redshift 
\citep{Kuk01}. However, other authors  \citep[e.g.][and references therein]{Fal08, SWJ08} found evidence that high-redshift hosts are brighter than 
can be explained by passive evolution or that the ratio of black-hole mass to spheroid 
mass is greater than that determined for the local universe \citep[e.g.][]{McL06, Peng06}.  The inconsistencies in the calculated relative strength of the presumed host
galaxy raises questions as to the overall validity of these
detections.  Also, it is generally found that powerful nuclear activity, typical of quasars, at {\it z} $\ga 0.4$ is predominantly 
associated with bulge-dominated galaxies, regardless of radio luminosity 
\citep{McL99, Dun03, McL04, Fal08}.

However, all these studies have neglected the potentially important 
contribution of scattered nuclear (quasar) light to the surrounding nebulosity.  In a luminous quasar, it is possible 
that scattering off free electrons and 
dust within the host galaxy will compete with, or possibly 
entirely dominate, the direct emission from the host galaxy itself.

The signature of the presence of significant scattered nuclear light, as 
witnessed by its polarization properties, is widespread in AGN.  Indeed, 
polarimetric measurements have brought about some of the greatest leaps in 
our physical understanding of these objects.  The most prominent example 
being the formulation of the unified scheme of AGNs \citep[e.g.][and 
references therein]{Ant93}. Subsequent studies have shown that many radio 
galaxies are surrounded by spatially resolved reflection nebulae (e.g., 
PKS 2152-69, \citeauthor{diSA88} \citeyear{diSA88} and Cygnus A, \citeauthor{Tad90} \citeyear{Tad90}), with the polarized 
region extending over several kpc \citep{diSA89}.

In Seyfert 2 galaxies (e.g., NGC 1068, \citeauthor{Cap95} \citeyear{Cap95}; Mrk 463, 
\citeauthor{Uom93} \citeyear{Uom93}; Mrk 3, \citeauthor{Kish02} \citeyear{Kish02}) essentially all the extended UV 
light is scattered.  In NGC 1068, the degree of polarization is as high as 
60\% in parts of the resolved scattering region. For the nearby Seyfert 1 
galaxy NGC 4151 it is possible to trace scattering both in the traditional 
scattering cone and also within the host galaxy coincident with the 
extended narrow line region \citep{Drap92, Rob04}.  Significant polarization due to scattering is observed to 
extend to a distance of greater than 1.4 kpc.

The greater majority of bright QSO, approximately 99\%, are only marginally polarized with a degree of polarization of less than 3\% and generally at the 1\% level \citep{SMA84}. However, polarization measurements by themselves carry little information on the level of scattering or where it occurs.  If the scattering region is unresolved, geometrical cancellation, i.e., summing of the polarization vectors oriented at different angles, as in the case of the centro-symmetric pattern, produces a null polarization.  The measured polarization level is only an indication of the irregularities in the scattering region and is therefore a {\em lower limit} to the scattered flux fraction.  This is true whether or not there is dilution, from a stellar component for example.  Thus, for a typical observed quasar polarization of 1\%, scattering will contribute 1\% of the total observed flux if the scattered light is intrinsically 100\% polarized.  If the AGN to host galaxy ratio is 100, then the scattered component is equal in strength to the host galaxy.  In real situations, an intrinsic polarization of 100\% is not possible, so in fact, the scattered flux contribution would have to exceed 1\% Êin order to maintain the same level of polarization.  If we were to take NGC 1068 as an example, with an integrated intrinsic 
polarization of 16\% \citep[e.g.][]{Young95}, then the scattered flux 
would account for 6.25\% of the total observed flux, i.e., over six times 
brighter than the host galaxy at an AGN to host galaxy ratio of 100.  Studies of the host galaxies of type 2 quasars, in which
the AGN is obscured in the direct view, indicates that scattered nuclear light
can be comparable to, or even dominate, the extended emission \citep{Zak06}.

In this Letter, we aim to determine whether scattering in the host galaxy is a potential concern to host parameterization in high-redshift observations.  In Section 2, we introduce the scattering model and in Section 3, we present the results of our modeling.  We discus these results in comparison to the observations in Section 4.

\section{The Scattering Model}

We take the scattering model of \citet{Young00} as a starting point.  
This model can incorporate any source and scattering geometry that can be 
described analytically. The model allows for scattering off free electrons 
and dust grains in the Rayleigh regime. For dust scattering, we assume a 
refractive index that matches that of silicates \citep{Simm82} and that 
the particle is a sphere with a diameter of 0.1\micron.

The simplest geometry to model, that has relevance to the problem at hand, 
is that of a sphere of scattering particles, be they dust particles or 
free electrons. It should be pointed out that the models presented here do 
not include any starlight from the host galaxy.  The aim is to show 
whether scattering alone can produce an extended illuminated region 
comparable to those detected about QSO nuclei.  In terms of the geometry 
of the scattering region, we will limit this case study to spheroid and 
disk regions.  The spheroid cases represent elliptical galaxies and the 
disk-like region will represent the later type spiral galaxies.  In 
addition, a combination of both the spheroid and disk regions can represent 
the early-type spiral and S0 galaxies.

In excess of 5000 individual model runs were executed, the ranges in input 
parameters are listed in Table 1.  To facilitate comparison with 
observations, it was then necessary to convolve the model output images 
with the point-spread function (PSF) of a specific instrument.  We chose to simulate images from the 
ACS and NICMOS cameras onboard the {\it Hubble Space Telescope} ({\it HST}) and thus, the images were binned at 
the respective pixel scales of 0.025 (for {\it z} = 0.4 galaxies for comparison 
with the observations of \citet{Floyd04}) and 0.043 arcsec (for higher {\it z} 
objects).  Model PSFs were constructed using the TINYTIM software using 
filters that corresponded to the rest wavelength range of \citet{Kuk01}.

The spheroid scattering region is defined in terms of an inner and outer 
scattering radius in the three principal geometrical axes and the 
inclination of the system {\it z}-axis to the line of sight.  This allows for a 
variety of spheroid shapes but by keeping the {\it x}- and {\it y}- radii equal the 
more specific cases of oblate and prolate spheroid regions can be studied.  
The number density of scattering 
particles is assumed to have a power-law distribution with radius, $n = n_o(r_o/r)^{\alpha}$, and the number density at the inner scattering radius, $n_o$, was chosen such that the 
scattering region remains optically thin.

The disk-like region is also defined in terms of an inner and outer radius 
in both the {\it x}- and {\it y}-axes with a variable vertical extent in the {\it z}-axis.  
Thus, the resultant geometry is that of a circular slab.

\section{Application to Radial Profiles}

The resultant radial profiles, for a simple spherical scattering region, 
are presented in Figure 1 for a redshift of {\it z} = 2 and convolved with the 
NICMOS PSF.  Here, the radial profiles are formed by calculating the 
cumulative flux in a circular aperture of increasing radius and then 
normalized to the flux in the largest aperture.  For the models presented 
in Figure 1, the scattering elements are dust particles and the optical 
depth to scattering is 0.5.  For the case with $\alpha = 0$, the 
uniform number density is $3 \times 10^{-8} $m$^{-3}$.  It is readily 
apparent, even with this simplistic arrangement, that there is significant 
extended flux, at least for the cases with $\alpha$ equal to 0 and 1.  
Indeed, with $\alpha = 0$ there is $\sim$ 2 times more flux at a 
radial distance of 0.5 arcsec than for the point-source PSF alone.

The relatively low observed polarization of most quasars (Section 1) is an 
important constraint on the scattering models.  The simple sphere of 
scattering particles produces a true centro-symmetric pattern of 
polarization vectors and on integration, for all values of aperture 
radius, the degree of polarization is zero.  A disk-like scattering 
region, with high optical depth to scattering (uniform electron number 
density of $10^{7} $m$^{-3}$ -- equatorial optical depth to scattering = 
0.5) at an inclination of 90\arcdeg gives an integrated degree of 
polarization of 1.3\% when the disk is thin (an aspect ratio of 
0.05), dropping to 0.8\% as the disk thickness increases (an aspect 
ratio of 0.2).

The ratio of the quasar flux to that of the ``host galaxy'' is an 
important parameter in the studies of high redshift AGN.  The simple 
sphere model in Figure 1, with $\alpha = 0$, results in a quasar to 
extended-flux ratio of 1.3.  Reducing the dust number density to produce 
an optical depth to scattering of 0.05 increases the ratio to 11.2.  For 
the disk-like scattering region at an inclination of 90\arcdeg and a 
uniform electron number density of $10^{7} $m$^{-3}$, the quasar to 
extended-flux ratio varies from 5.0 (aspect ratio of 0.05), through 2.6 
(aspect ratio of 0.1) to 1.5 for an aspect ratio of 0.2.  The average 
quasar to ``host'' ratio for the quasars at {\it z} = 2 listed in \citet{Kuk01} 
is 3.6.  In Figure 2, we present the quasar/extended-emission flux ratio as a 
function of optical depth for various spheroid and disk models with 
$\alpha$ = 0, showing when the extended-emission levels become comparable 
to those observed by \citet{Kuk01}.  With an optical depth of only 0.1, 
scattering can easily produce $\ga$50\% of the detected extended flux.  
Such small optical depths and the implied absorption depths are consistent 
with the average extinction suffered by quasars \citep[e.g.][]{Hop04}.

\section{Discussion} 

A quasar point source embedded in a geometry similar to either disk or spheroid galaxy types 
can produce significant extended flux.  The question remains is this 
likely to make a significant contribution to the actual observed extended 
flux?  \citet{Kuk01} illustrated their observations in similar encircled 
energy verses radius plots and our Figure 1 can be compared to their Figure 4.  
It is readily apparent that scattering in a spherical distribution of 
scattering elements can produce very similar results to the observations 
reported by these authors.  Indeed, the models presented in our Figure 1 
exceed the extended flux levels reported.

If the scattering particles are free electrons, rather than dust grains as in the models shown in Figure 1, the number densities required to produce extended-flux levels comparable to those observed are of the order of $10^{6}$--$10^{7} $m$^{-3}$.  This is a similar order of magnitude to the average 
warm interstellar medium (ISM) in our Galaxy at $10^{6}$m$^{-3}$ 
\citep[][and references therein]{Fer01}.  A quasar is likely to ionize the 
tenuous component of the ISM to a distance of several kpc \citep{Net04} 
and models of {\it z} = 6 galaxies \citep{Yu05} suggest that quasar host 
galaxies could be nearly totally ionized.  The uniform dust number density 
in the models illustrated in Figure 1 is of the order of that expected for a 
normal galaxy based on the $\rho_d/\rho_H$ ratios for our Galaxy 
\citep{Cong05}.  Host galaxies of quasars are in general bluer than 
inactive galaxies \citep{Canal06, Canal07, Jahn07}, a fact that is used to infer a high rate of star formation, but could be explained by the wavelength dependence of dust scattering.  Dependent on the medium, constraints to the magnitude of the scattered component may prove observable.  As examples, in the case of electron scattering the presence and strength of recombination lines;\footnote{Only the case with uniform number density of $10^{7} $m$^{-3}$ and an outer radius of 20 kpc will produce a significant line luminosity} for dust scattering the color of the resultant extended emission.  Such detail will be the subject of a forthcoming modeling paper.

In the local universe, the ratio of quasar to the host galaxy luminosity can be as 
large as 2 orders of magnitude.  For example, for a selection of quasars with 
an average redshift of {\it z} = 0.17 this ratio is $\sim$25 \citep{Wolf08}.  However, studies of 0.4 $\la$ {\it z} $\la$ 3 quasars \citep{Kuk01, Floyd04, Fal08} frequently find ratios 
between 1 and 10, sometimes less than 1.  Given that quasars were significantly more luminous at earlier epochs, this would seem to require strong evolution of the host galaxy, if scattering
of nuclear light is insignificant. In fact, the
inferred decrease in host luminosity from $z\sim 1$
to $z = 0$ is significantly smaller than the decrease
in quasar luminosity \citep{Kuk01}. 
However, if scattering dominates the extended emission, as seems to be the case in at least some
type 2 quasars \citep{Zak06}, then  relatively small decreases in the scattering optical depth
can produce large increases in the quasar/host luminosity ratio. For example, it can be seen from Figure 2
that the scattered light quasar/``host'' ratio is $\sim 2$, for optical depths $\sim 0.1 - 0.3$, but
increases rapidly for optical depths $< 0.1$. We suggest, therefore, that the apparent
evolution of the quasar/host luminosity ratio arises because host galaxies at $z \sim 0$ contain
less scattering material than their high-redshift counterparts. This might arise as a result of
star formation, or because the less luminous quasar nucleus photoionizes 
a smaller fraction of the ISM.  As the observed extended flux is a combination of scattered quasar and stellar emission, stellar features would still be present in the ``host'' spectra, albeit diluted by scattered light.  Given the evolution of the quasar luminosity function with redshift, it is expected that these features would become more prominent in more recent epochs.  The equivalent width of the stellar features may therefore provide a useful constraint on the strength of the scattered component.

RLQ are, on average, more luminous than their RLQ counterparts, by approximately 1 mag at $z = 2$ \citep{Kuk01}.  These authors also find that, on average, the hosts of radio-loud objects are the more luminous by 1.5 mag at the same epoch.  If scattering contributes significantly to the observed extended emission then, as already stated, the perceived luminosity of the host will be loosely tied to that of the quasar.  In this case, the finding of brighter radio-loud hosts is possibly a natural consequence of the average intrinsic luminosity difference between RLQ and RQQ.

In the local universe, radio-loud objects are predominantly 
associated with elliptical galaxies and radio-quiet objects with disk galaxies.  
Studies of host galaxies at high redshift have concluded that quasars generally inhabit elliptical galaxies regardless of radio power \citep[e.g.][]{Kuk01}. However, if scattering is of importance, it is possible that the observed radial profiles may be influenced as much by the electron or dust distribution as the stellar light distribution, raising doubt as to the actual galaxy type.  This possibility will be explored in greater depth in a forthcoming paper.

Scattering in the host galaxy around quasars, whilst providing extended 
flux, is merely a reflection of the quasar itself.  Thus, any changes in 
the luminosity of the AGN will propagate out as a moving light echo.  
\citet{Benn08} present observations of several relatively low redshift ($z 
= 0.14-0.21$) quasars that show shells and other structures that these 
authors attribute to merger remnants.  Whilst we have no direct 
polarimetry evidence to refute that assertion, we speculate that some of 
the structures observed in these objects might actually be light echoes 
resulting from past high-luminosity outbursts from the quasar.

It has been concluded \citep{Kuk01, Floyd04, Fal08} that the host galaxies 
of quasars are essentially fully formed by $z = 2$, or earlier, and then 
undergo passive evolution to the current epoch.  Other authors 
\citep{Fal08, SWJ08} determined that the host galaxies of the quasars were 
a factor of 3 to 5 times more luminous than allowed by passive evolution 
alone.  However, scattering in the host
galaxy can easily contribute a significant fraction of the extended flux
detected around quasars at high-{\it z}, leading to overestimates of the
host luminosities and masses, which in turn may give rise to a misleading picture of evolutionary
trends.

\section{Summary and Conclusions}

We have presented models that indicate that scattering in the ISM of the host galaxy around quasars could significantly alter the characterization of the host themselves.  Whilst scattering may be no more that a nuisance effect at low-{\it z}, with the increase in the quasar luminosity with increasing redshift this may represent a major problem for host galaxy studies.  Indeed, it is possible that the stellar emission from the actual host has not been detected for high-redshift quasars.  For electron scattering, a number density of the same order as that in the Milky Way, together with a high ionization degree ($\sim$50\%) is required.  If the scattering medium is predominantly dust grains and uniformly distributed, then a dust number density similar to that in our Galaxy can produce the required extended flux without the requirement for a high degree of ionization.

Such scattering naturally explains the trend for more luminous ``hosts'' around the brighter RLQ compared to those of the RQQ, since the extended flux is directly related to the luminosity of the quasar.  If the real host has not been detected, or at least over estimated in terms of its luminosity, then it is not possible to determine evolutionary trends with any degree of accuracy.  One obvious observational test of the idea presented in this Letter is to repeat the high spatial resolution observations of high-redshift quasars made by other authors but include a polarimetry module to determine the polarization parameters of the potential scattered flux.  High spatial resolution is essential to this type of observation to avoid smearing out the polarization and rendering it undetectable.

\acknowledgments

We thank the anonymous referee for comments and suggestions that have greatly improved the clarity of this Letter.

\clearpage

\begin{figure}
\epsscale{0.80}
\plotone{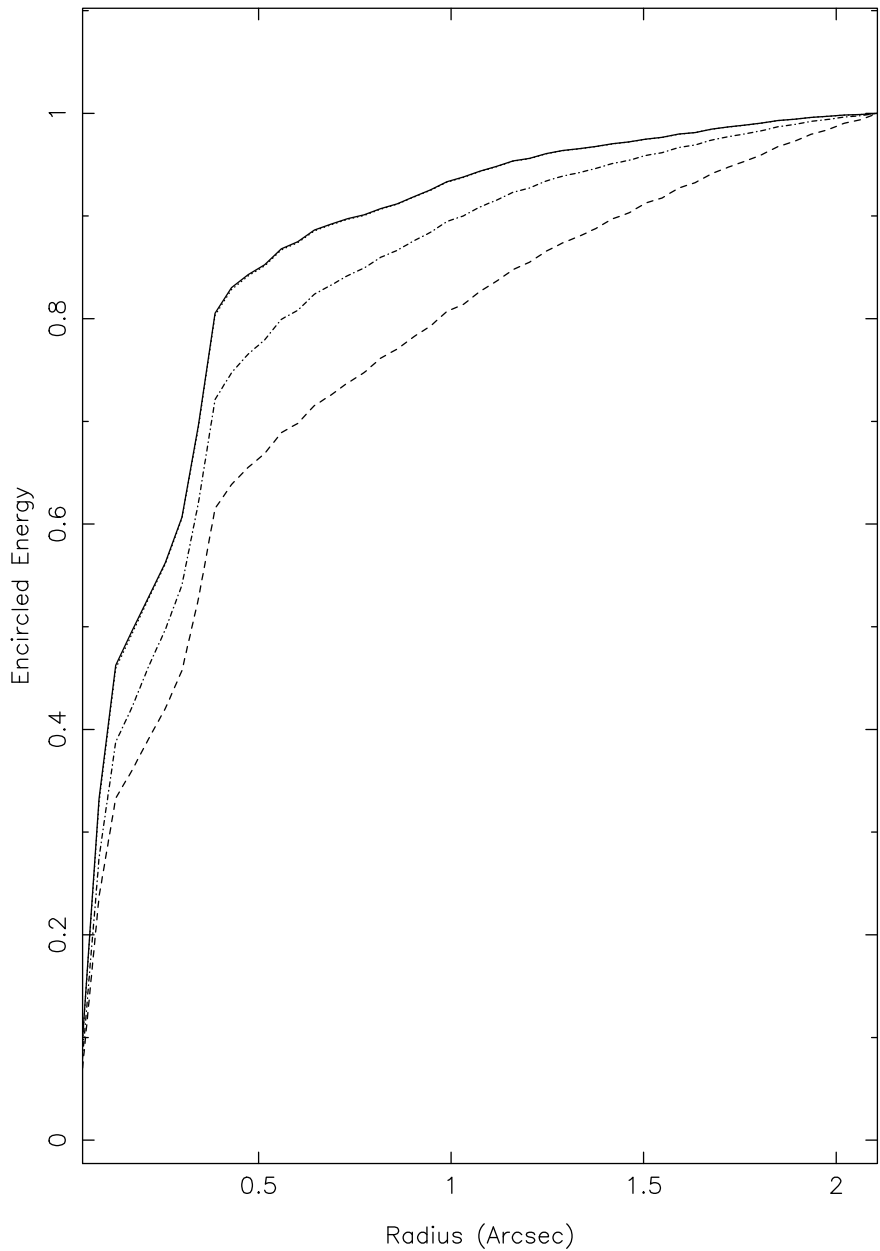}
\caption{Azimuthally averaged radial profile for the NICMOS PSF (solid line) and three models representing scattering in a sphere of dust grains.  The geometrical parameters of the sphere are described in the main text, but the exponent of the number density power-law is 0 (dashed line), 1 (dot-dashed line), and 2 (dotted line) with the number density at the inner scattering radius set to give equal optical depth to scattering in each case.  As plotted here, the profile for $\alpha$ = 2 is indistinguishable from the PSF profile.\label{fig1}}
\end{figure}

\clearpage

\begin{figure}
\epsscale{0.80}
\plotone{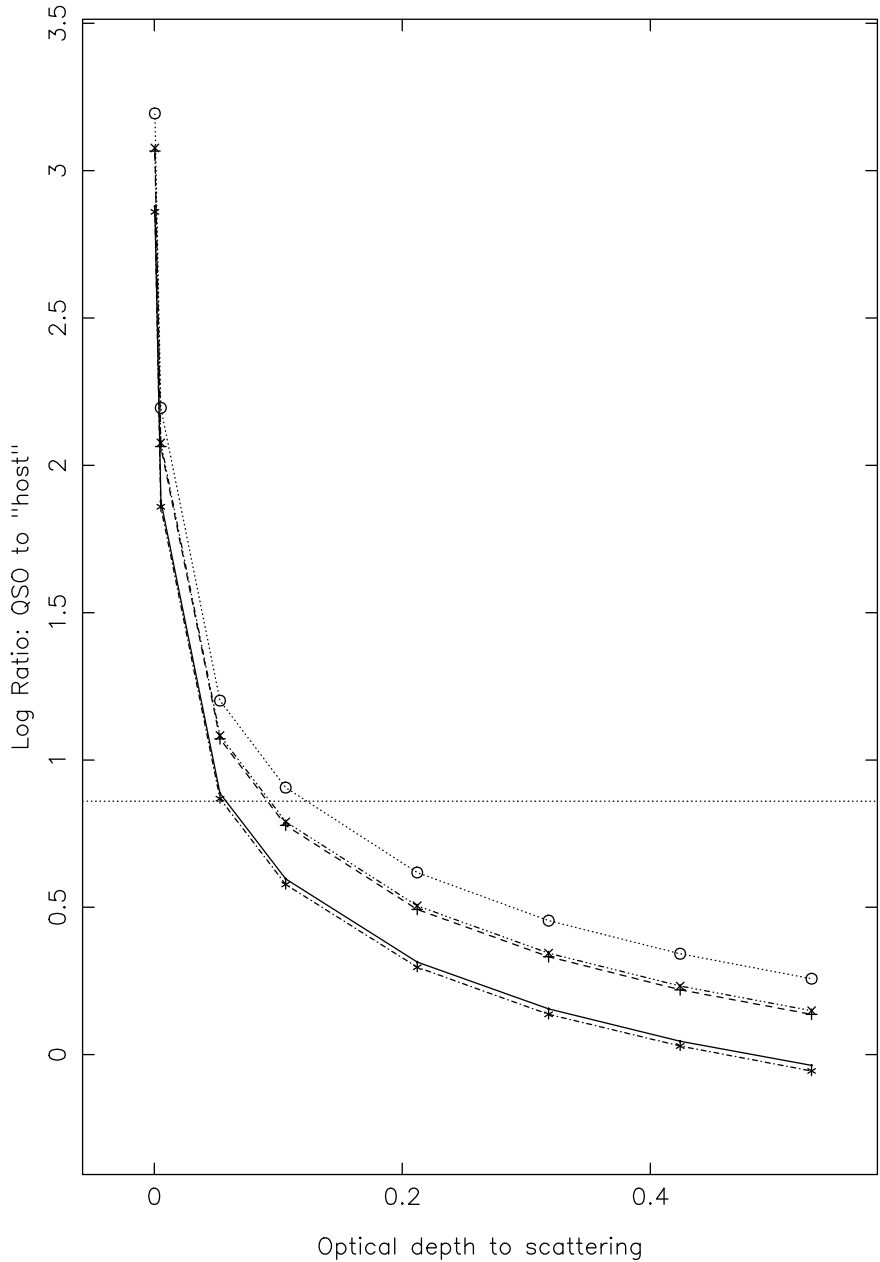}
\caption{The ratio of the quasar flux to that of the extended flux for varying optical depth to scattering.  Models for several scattering geometries are plotted: spherical (solid line), 
oblate (dashed line), and prolate (dot-dashed line) spheroids at zero inclination and
with outer scattering radii along the {\it z}-axis of 0.3 and 1.2$\times$ the radius in the $x-y$ plane, respectively; disk with an aspect ratio of 0.2, inclinations of
0\arcdeg (dotted line) and 60\arcdeg (triple-dot--dashed line).  Other parameters are as described in the main text.  The horizontal dotted line represents a quasar to host ratio with an extended flux equal to half the average of the \citet{Kuk01} galaxies.\label{fig2}}
\end{figure}

\clearpage

\begin{deluxetable}{lllrr}
\tabletypesize{\scriptsize}
\tablecaption{Parameters used in the modeling.}
\tablewidth{0pt}
\tablehead{\colhead{Parameter} & & & \colhead{Values} & \colhead{Units}}
\startdata
Inclination & & & 0, 30, 60, 90 & \arcdeg \\
$\alpha$ ($n = n_o(r_o/r)^{\alpha}$) & & & 0, 1, 2 & - \\
$n_o$ (electrons; $\alpha$ = 0) & & & $1 \times 10^{4}-1 \times 10^{7}$ & m$^{-3}$ \\
$n_o$ (dust; $\alpha$ = 0) & & & $1 \times 10^{-10}-1 \times 10^{-6}$ & m$^{-3}$ \\
Distance & & & 0.3, 1, 2, 4 & $z$ \\
$r_o$ & & & 10 & pc \\
$r$ & Simple sphere & & $2.5-20$ & kpc \\
 & Spheroid & $x, y$: & $2.5-20$ & kpc \\
 & & $z$: & $8.7-80$ & \\
 & Disk & $x, y$: & $2.5-20$ & kpc \\
 & & $z$ (height): & $1-4$ & \\
\enddata
\end{deluxetable}


\begin{thebibliography}{}
\bibitem[Antonucci(1993)]{Ant93} Antonucci, R. R. J.  1993, \araa, 31, 473
\bibitem[Aretxaga et al.(1998)]{ATB98} Aretxaga, I., Terlevich, R. J. \& Boyle, B. J. 1998, \mnras, 296, 643

\bibitem[Bennert et al.(2008)]{Benn08} Bennert, N., Canalizo, G., 
Jungwiert, B., Stockton, A., Schweizer, F., Peng, C. Y. \& Lacy, M.  2008, 
\apj, 677, 846
\bibitem[Canalizo et al.(2007)]{Canal07} Canalizo, G., Bennert, N., Jungwiert, B., Stockton, A., Scheizer, F., Lacy, M. \& Peng, C.  2007, \apj, 669, 801
\bibitem[Canalizo et al.(2006)]{Canal06} Canalizo, G., Stockton, A., Brotherton, M. S. \& Lacy, M.  2006, NewAR, 50, 650
\bibitem[Capetti et al.(1995)]{Cap95} Capetti, A., Axon, D. J., Macchetto, F., Sparks, W. B. \& Boksenberg, A.  1995, \apj, 446, 155
\bibitem[Congui et al.(2005)]{Cong05} Congui, E., Geminale, A., Barbaro, G. \& Mazzei, P.  2005, J. Phys.: Conf. Ser., 6, 161
\bibitem[di Serego-Alighieri et al.(1988)]{diSA88} di Serego-Alighieri, S., Courvoisier, T. J. -L., Fosbury, R. A. E., Tadhunter, C. N. \& Binette, L.  1988, \nat, 334, 591
\bibitem[di Serego-Alighieri et al.(1989)]{diSA89} di Serego-Alighieri, S., Fosbury, R. A. E., Tadhunter, C. N. \& Quinn, P. J.  1989, \nat, 341, 307
\bibitem[Draper et al.(1992)]{Drap92} Draper, P. W., Gledhill, T. M., Scarrott, S. M. \& Tadhunter, C. N.  1992, \mnras, 257, 309
\bibitem[Dunlop et al.(2003)]{Dun03} Dunlop, J. S., McLure, R. J., Kukula, M. J., Baum, S. A., O' Dea, C. P. \& Hughes, D. H.  2003, \mnras, 340, 1095
\bibitem[Falomo et al.(2008)]{Fal08} Falomo, R., Treves, A., Kotilainen, J. K., Scarpa, R. \& Uslenghi, M.  2008, \apj, 673, 694
\bibitem[Ferri\`{e}re(2001)]{Fer01} Ferri\`{e}re, K. M. 2001, Rev. Mod. 
Phys., 73, 1031
\bibitem[Floyd et al.(2004)]{Floyd04} Floyd, D. J. E., Kukula, M. J., Dunlop, J. S., McLure, R. S., Miller, L., Percival, W. J., Baum, S. A. \& O' Dea, C. P.  2004, \mnras, 355, 196
\bibitem[Hopkins et al.(2004)]{Hop04} Hopkins, P. F., et al. 2004, \aj, 128, 1112
\bibitem[Jahnke et al.(2007)]{Jahn07} Jahnke, K., Wisotzki, L., Courbin, F. \& Letawe, G.  2007, \mnras, 378, 23
\bibitem[Kishimoto et al.(2002)]{Kish02} Kishimoto, M., Kay, L. E., Antonucci, R., Hurt, T. W., Cohen, R. D. \& Krolik J. H.  2002, \apj, 565, 155
\bibitem[Kukula et al.(2001)]{Kuk01} Kukula, M. J., Dunlop, J. S., McLure, R. J., Miller, L., Percival, W. J., Baum, S. A. \& O' Dea, C. P., 2001, \mnras, 326, 1533
\bibitem[McLure et al.(2006)]{McL06} McLure, R. J., Jarvis, M. J., Targett, T. A., Dunlop, J. S. \& Best, P. N.  2006, \mnras, 368, 1395
\bibitem[McLure et al.(1999)]{McL99} McLure, R. J., Kukula, M. J., Dunlop, J. S., Baum, S. A., O' Dea, C. P. \& Hughes D. H.  1999, \mnras, 308, 377
\bibitem[McLure et al.(2004)]{McL04} McLure, R. J., Willott, C. J., Jarvis, M. J., Rawlings, S., Hill, G. J., Mitchell, E., Dunlop, J. S. \& Wold, M.  2004, \mnras, 351, 347
\bibitem[Netzer et al.(2004)]{Net04} Netzer, H., Shemmer, O., Maiolino, R., Oliva, E., Croom, S., Corbett, E. \& di Fabrizo, L. 2004, \apj, 614, 558
\bibitem[Peng et al.(2006)]{Peng06} Peng, C. Y., Impey, C. D., Rix, H-W., Kochanek, C. S., Keeton, C. R., Falco, E. E., Leh$\acute{a}$r, J. \& McLeod, B. A.  2006, \apj, 649, 616
\bibitem[Robinson et al.(1994)]{Rob04} Robinson, A., et al. 1994, \aap, 291, 351
\bibitem[Schramm et al.(2008)]{SWJ08} Schramm, M., Wisotzki, L. \& Jahnke, K.  2008, \aap, 478, 311
\bibitem[Simmons(1982)]{Simm82} Simmons, J. F. L.  1982, \mnras, 200, 91
\bibitem[Stockman et al.(1984)]{SMA84} Stockman, H. S., Moore, R. L. \& Angel, J. R. P.  1984, \apj, 279, 485
\bibitem[Tadhunter et al.(1990)]{Tad90} Tadhunter, C. N., Scarrott, S. M. \& Rolph, C. D.  1990, \mnras, 246, 163
\bibitem[Uomoto et al.(1993)]{Uom93} Uomoto, A., Caganoff, S., Ford, H. C., Rosenblatt, E. I., Anotnucci, R. R. J., Evans, I. N. \& Cohen, R. D. 1993, \aj, 105, 1308
\bibitem[Wolf \& Sheinis(2008)]{Wolf08} Wolf, M. J. \& Sheinis, A. I. 2008, \aj, 136, 1587
\bibitem[Young et al.(1995)]{Young95} Young, S., Hough, J. H., Axon, D. J., Bailey, J. A. \& Ward, M. J.  1995, \mnras, 272, 513
\bibitem[Young(2000)]{Young00} Young, S.  2000, \mnras, 312, 567
\bibitem[Yu \& Lu(2005)]{Yu05} Yu, Q. \& Lu, Y. 2005, \apj,, 620, 31
\bibitem[Zakamska et al.(2006)]{Zak06} Zakamska, N. L., Strauss, M. A., Krolik, J. H., Ridgway, S. E., Schmidt, G. D., Smith, P. S., Heckman, T. \& Schneider, D. P. 2006, \aj, 132, 1496

\end{thebibliography}
\end{document}